\documentclass[aps,prb,amsmath, amssymb, english, twocolumn,reprint,floatfix, superscriptaddress]{revtex4-1}

\usepackage{bbm}
\usepackage{hhline}
\usepackage{multirow}

\usepackage{tikz}
\usepackage{pgfplots}
\usepackage{microtype}
\usepackage{tkz-euclide}
\usepackage{epstopdf}
\usepackage{epsfig} 
\usepackage{amsmath}

\usepackage{hyperref}
\hypersetup{
    colorlinks,
    citecolor=blue,
    filecolor=blue,
    urlcolor=blue}

\newcommand{\bR}{\mathbf{R}}

\newcommand{\cc}{c^{\ }}
\newcommand{\bq}{\mathbf{q}}

\newcommand{\bk}{\mathbf{k}}

\newcommand{\cd}{c^\dag}

\newcommand{\psid}{\psi^\dag}

\newcommand{\phid}{\phi^\dag}

\allowdisplaybreaks

\begin{document}

\title{Dynamical vertex approximation for the attractive Hubbard model}
\author{Lorenzo Del Re} 
\affiliation{International School for
  Advanced Studies (SISSA), Via Bonomea
  265, I-34136 Trieste, Italy}
\affiliation{Institute for Solid State Physics, TU Wien, 1040 Vienna, Austria}
\author{Massimo Capone}
\affiliation{International School for
  Advanced Studies (SISSA), Via Bonomea
  265, I-34136 Trieste, Italy}
\author{Alessandro Toschi} 
\affiliation{Institute for Solid State Physics, TU Wien, 1040 Vienna, Austria}

\date{\today} 


\begin{abstract}
In this work, we adapt the formalism of the dynamical vertex approximation (D$\Gamma$A), a diagrammatic approach including many-body correlations  beyond  the dynamical mean-field theory, to the case of {\sl attractive} onsite interactions. 
We start by exploiting the ladder approximation of the D$\Gamma$A scheme, in order to derive the corresponding equations for the non-local self-energy and vertex functions of the attractive Hubbard model. 
Second, we prove the validity of our derivation by showing that the results obtained in the particle-hole symmetric case fully preserve the exact mapping between the attractive and the repulsive models. It will be shown, how this property can be related to the structure of the ladders,  which makes   our derivation applicable for any approximation scheme based on ladder diagrams.  
Finally, we apply our D$\Gamma$A algorithm to the attractive Hubbard model in three dimensions, for different fillings and interaction values. Specifically, we focus on the parameters region in the proximity of the second-order transition to the superconducting and charge-density wave phases, respectively, and calculate (i) their phase-diagrams, (ii)  their critical behavior, as well as (iii) the effects of the strong non-local correlations on the single-particle properties.
\end{abstract}
\maketitle
\section{Introduction}

The theoretical description of strongly correlated fermionic systems represents one of the main challenges in condensed matter physics. This field has been defined by the study of the two-dimensional Hubbard model as the prototype simple model to understand high-temperature superconductivity in copper oxides and it has flourished thanks to the discovery of a variety of materials with strongly correlated valence electrons displaying remarkable functional properties. 
The importance of the physics emerging in these systems and the complexity of actual corrrelated materials  have motivated the quest for reproducing, in a much more controlled way, the fundamental microscopic mechanisms at work. This is possible, e.g., by trapping systems of cold atoms in optical lattices induced through laser interference \cite{doi:10.1146/annurev-conmatphys-070909-104059}. 

Among the plethora of interesting phenomena associated with strong correlations, some of the most difficult problems arise when the electrons are confined in low spatial dimensions  (as in anisotropic layered transition-metal-oxides) or in the proximity of phase transitions of different kind. 
In these regimes, one finds that the electron correlations appear simultaneously in space and time coordinates or, in a more mathematical language, the self-energy depends both on momentum and on frequency and those dependencies are intertwined.

Indeed, the effort to develop theoretical methods able to treat strong correlations has produced, among the others,  a rigorous non-perturbative theoretical treatment of quantum local correlations, through the popular and succesful dynamical mean-field theory (DMFT) \cite{georges1996dynamical}, where the self-energy is momentum-independent, but it has a rich frequency dependence.  DMFT, however, becomes exact\cite{georges1996dynamical} only for lattices with large coordination number (or, equivalently, in the limit of high dimensions) which obviously limits severely its application to low-dimensional systems and close to phase transitions where the electronic self-energy is expected to be strongly modified by non-local fluctuations. 

The efforts to introduce non-local effects starting from DMFT has led to cluster extensions\cite{MaierCluster} or diagrammatic expansions\cite{rohringer2018diagrammatic} of the method. The latter ones use the local two particle vertex functions\cite{RohringerValli2012,schaefer2013} of DMFT (or, more precisely, of its auxiliary Anderson impurity model, AIM) as a building block for new diagrammatic expansions. For instance, the dual fermion (DF)\cite{Rubtsov2008} and the dynamical vertex approximation (D$\Gamma$A)\cite{Toschi2007} schemes use the full two-particle local vertex function $\mathcal{F}$ and the two-particle irreducible local vertex functions, respectively, to include non-local correlation effects on top of DMFT,  through ladder or parquet diagrammatic resummations. 

These approaches have been so far applied essentially only to the cases of repulsive (and , often, local or short range) electronic interactions . On the other hand,  since the dynamical mean-field theory has been also successfully used for studying systems with \emph{attractive} interactions \cite{metznerAttractive, capone2002,Toschi2005-1, toschi2005, garg2005, privitera2010,koga2011, amaricci2014, agnese2016} and  the Feynman diagrammatics  represents an highly flexible tool, no conceptual obstacle should prevent the applicability of the diagrammatic approaches beyond DMFT to systems with attractive interactions, which have so far been considered only in Ref. \cite{RUBTSOV2012}, where DB has been applied to an extended attractive Hubbard model \cite{van2017extended}.

We also discuss here the relation between the D$\Gamma$A and the purely numerical lattice Quantum Monte Carlo (QMC)\cite{Hirsch}, that for the attractive model can be regarded as numerically exact, but it can only be implemented in finite lattices which require a finite-size scaling to estimate the thermodynamic limit.
The situation is somewhat complementary in  D$\Gamma$A, at least in the ladder version used in this work. The approach introduces approximations, but it is free from finite-size limitations. Furthermore, in the ladder-D$\Gamma$A we have a clear information about which are the fluctuations (bosonic modes) that enters in the self-energy through Schwinger-Dyson equation 
with the possibility of performing fluctuations diagnostic \cite{Gunnarsson2015}. 

The extension of diagrammatic approaches to the attractive Hubbard model is directly relevant to describe systems of cold atoms in optical lattices
where the interaction is tuned to attractive values exploiting Feshbach resonances\cite{inouye1998observation,RevModPhys.82.1225}  and the coupling is tuned from weak to strong giving rise to the  BCS-BEC crossover\cite{PhysRevLett.91.250401,PhysRevLett.92.040403,Chin1128,PhysRevLett.93.050401,PhysRevLett.92.150402,zwierlein2005vortices,PhysRevLett.92.120401}. In this framework, significant corrections with respect to static mean-field descriptions are expected in particular at intermediate interaction values and for low-dimensional systems. Moreover, the approach can be used to study mass-imbalanced\cite{PhysRevB.76.104517} or spin-imbalanced\cite{PhysRevLett.101.236405} mixtures beyond DMFT.
While an attractive Hubbard model is hardly applicable to concrete materials, its understanding can improve our fundamental knowledge of the properties of the superconducting phases in the regime of intermediate coupling where simple analytical approaches can not be applied.
Finally, a wider application of the diagrammatic extensions of DMFT to treat different physical situations from those considered hitherto (mostly systems with predominance of spin and charge fluctuations) will define, much more precisely, the strengths and the limitations of these methods -- a very important issue in the context of non-perturbative schemes, for which the number of possible benchmarks is severely limited by the complexity of the physical systems to describe. 

In this paper, we show how one of the diagrammatic extensions of DMFT, the Dynamical Vertex Approximation (D$\Gamma$A), can be applied to treat systems with on-site {\sl attractive} interactions.  In particular,  we will consider its more commonly used implementation (which exploits a ladder approximation and Moriya corrections of the physical propagators). In this framework, we will show explicitly  how the corresponding  D$\Gamma$A equations can be generalized to treat the case of the attractive Hubbard model. The validity of our procedure will be proved, by demonstrating through analytical and numerical calculations that, for the particle-hole symmetric case, the exact properties related to the mapping between the repulsive and the attractive local (Hubbard) interaction is fully preserved by the ladder D$\Gamma$A schemes: The resulting D$\Gamma$A self-energy $\Sigma({\bf k}, i\nu)$ is  invariant under the sign change of the interaction $U \leftrightarrow  -U$.

We will also discuss, how this result is related to the general properties of the ladder resummation of diagrams, which makes it  applicable to all diagrammatic schemes built on such resummations, such as -in the context of the diagrammatic extensions of DMFT-  to DF\cite{Rubtsov2008}, DMFT + FLEX\cite{kitatani2015}, GW + DMFT\cite{AyralGW2103}, TRILEX\cite{AyralTRILEX2015}, or the 1PI\cite{Rohringer2013} approach.

Finally, the ladder D$\Gamma$A equations, derived in this paper, will be applied to the attractive Hubbard model  on a cubic lattice for different hole-doping levels and interactions,  to analyze the progressive differentiation between the transitions to the s-wave superconducting and the charge density wave (CDW) phases, degenerate only at half-filling. In particular,  (i) we will compute the corresponding phase diagram, for the most challenging condition of intermediate coupling, by determining the reduction of the ordering temperatures induced by non-local superconducting and charge correlations; (ii) we will then investigate the  critical properties of the two distinct transitions, as described by the D$\Gamma$A; (iii) we will, eventually,  analyze the effects of the underlying spatial and temporal fluctuations on the single-particle properties. 

Our formalism can be also applied to the two-dimensional model, where superconductivity appears through a more elusive Berezinskii-Kosterlitz-Thouless  transition, whose description requires a proper inclusion of phase fluctuations of the order parameter. This would require a further extension of our algorithm which goes beyond the aim of this paper. However, we expect the present algorithm to give sufficently accurate results far above the superconducting/superfluid critical temperature, i.e. in a parameter region where many-body pairing has been recently observed experimentally\cite{Murthyeaan5950}.

The paper is organized as follows:  In Sec.(\ref{sec:II}), we 
focus on the formal aspects of our work, introducing the general ideas behind the ladder-D$\Gamma$A method  extending the formalism 
in order to treat models with contact attractive interactions.
Part of the corresponding analytical derivations including the proof of the particle-hole symmetry consistency of the new scheme is reported in Appendix \ref{Proof}, while numerical checks of the consistency are shown at the end of the section.

In Sec.(\ref{sec:III}), we present
the results obtained applying the extended ladder-D$\Gamma$A algorithm to the 3D attractive Hubbard model away from half-filling. In particular, focusing on  
the region of parameters close to second order phase transitions to SC and CDW states, respectively, we analyze the effect of non-local spatial fluctuations in (i) reducing the ordering temperatures ($T_c$), (ii) modifying the critical behavior at the phase transitions, (iii) reducing the coherent motion of electrons.
Finally a brief summary is provided in Sec.(\ref{sec:concl}).

\section{D$\Gamma$A  for the attractive case}\label{sec:II}

In this section, we present the derivation of the equations for the electronic self-energy and susceptibilities for a single-orbital model with contact attractive interaction within D$\Gamma$A.
The model reads:
\begin{equation}
H =  -t\sum_{\langle ij\rangle \sigma }c_{i\sigma }^{\dagger }c_{j\sigma
}+{U}\sum_{i}\hat{n}_{i\uparrow }\hat{n}_{i\downarrow } -\mu\sum_{i\sigma}\hat{n}_{i\sigma},
\label{eq:hub}
\end{equation}
where  $c_{i\sigma }^{\dagger }$($c_{i\sigma }$) creates
(annihilates) an electron with spin $\sigma $ on site $i$, $\hat{n}_{i\sigma
}\!=\!c_{i\sigma }^{\dagger }c_{i\sigma }$,  $t$ denotes the hopping amplitude between nearest-neighbors, $\mu$ is the chemical potential and $U$ the on-site interaction, which takes negative values in the case of the Attractive Hubbard Model (AHM).
Following the convention of previous D$\Gamma$A publications, we use $D = 2\sqrt{6}t$ as the energy unit\ \  \cite{Rohringer2011}.

While D$\Gamma$A equations, at different approximation level, have been derived for the case of the repulsive ( $U> 0 $) Hubbard model,  in the course of several dedicated studies\cite{Toschi2007,Katanin2009, Toschi2011, Rohringer2016,  Galler2016, rohringer2018diagrammatic}, no explicit derivation has been presented so far for $U <0$. We notice that, in general terms, a formally simple change of sign of the interaction does not reflect in a straightforward generalization of the theoretical treatment. More in detail, extending a particular approximation scheme  may require special care if the original approximation was strongly based on the physics of the model (for example assuming that some kind of correlations are more important than others). 

In the specific case of the dynamical vertex approximation, it is rather clear that the generalization to the attractive case of the formalism of the full D$\Gamma$A scheme, based on an exact solution of the parquet equations, would be completely straightforward.
In fact, the parquet scheme treats all channels on equal footing, and thus, the D$\Gamma$A formalism can be used essentially unaltered  for repulsive and attractive interaction: In this scheme, one will always need to extract the local fully 2PI vertex from the corresponding DMFT calculations (either for the attractive or the repulsive model), and use it as input for the solver of the parquet equations on the chosen lattice.

The generalization procedure is less straightforward, instead, in the case of interest for this work, i.e., for the ladder D$\Gamma$A approximation.
In fact, the ladder D$\Gamma$A is based on the indentification of one (or more) dominant scattering channel(s), to which the non-local treatment of D$\Gamma$A will be specialized. 
In this situation, the extension of the equation will  clearly require a change in the selection of the non-local scattering channels, in order to have consistent results for the two models, as we shall discuss in details in the following.

In order to test the consistency of the ladder D$\Gamma$A theories for the attractive and repulsive Hubbard model we will exploit the unitary  transformation which maps the two model one onto the other, also known as Shiba transformation \cite{ShibaT}. Verifying whether the mapping is preserved for the ladder D$\Gamma$A equations of the two cases, will provide a solid testbed for the actual equivalence of the corresponding derivations.

\subsection{Ladder-D$\Gamma$A equations for the AHM}\label{sec:ladder-dga}

To understand the derivation of the ladder D$\Gamma$A equations for the AHM, it is useful to have in mind the corresponding derivation for the repulsive model. As the latter has been already reported in several studies\cite{Toschi2007,Held2008,Katanin2009,Rohringer2016} and in a recent review \cite{rohringer2018diagrammatic}, to which we refer the reader for details,   here we will only recapitulate the essential steps, focusing on the specific  modifications necessary due to the presence of an attraction.
\begin{figure}
\includegraphics[width = \columnwidth]{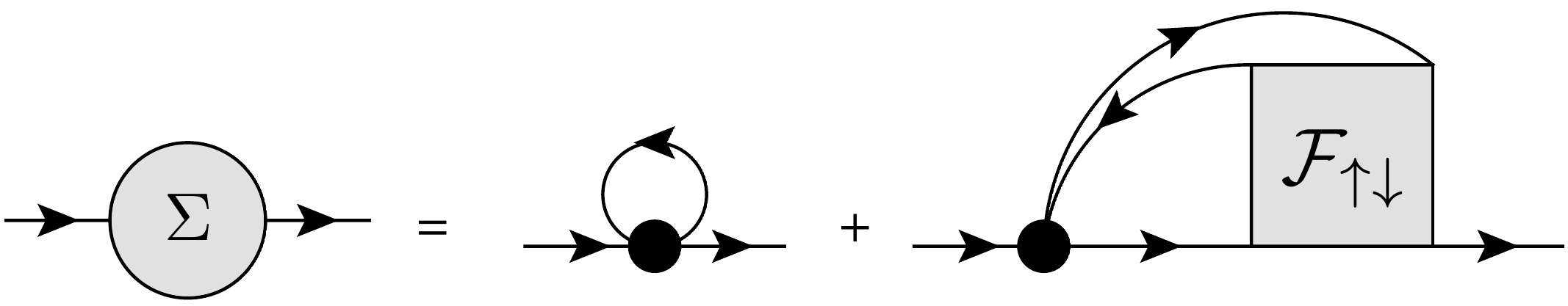}
\caption{Schwinger-Dyson equation for the self-energy. Arrowed lines refer to single-particle fermionic propagators, while thick dots to the bare contact interaction.}
\label{SD:eq}
\end{figure}

The D$\Gamma$A  relates the electronic self-energy $\Sigma$ to the full scattering amplitude $\mathcal{F}$ exploiting the exact Schwinger Dyson equation (see the schematic representation in Fig.(\ref{SD:eq}))
\begin{eqnarray}\label{eq:SDE}
&&\Sigma(k)-{Un\over 2}            \\
&&=-{U\over{\mathcal V}^2 } \sum_{k',q} {\mathcal F}_{\uparrow \downarrow}(k,k',q)    
G(k')G(k'+q)G(k+q)  \nonumber
\end{eqnarray}
where $G = G_\uparrow=G_\downarrow$ is the spin-independent single-particle Green's function,  $\mathcal{V}\equiv V /T$ ,  $k =(\bk,\nu)$, $q =(\bq,\omega)$ where $\nu$ and $\omega$ are the fermionic and bosonic Matsubara frequencies respectively.  We recall that    the single particle Green's function $G(\mathbf{k}, \nu)$  and the full scattering amplitude $\mathcal{F}$ are  related to the connected part of the two-particle Green's function $G^{(2)}_{c}$ through\cite{negele1988quantum}:
\begin{equation}\label{G2:connected}
G^{(2)}_{c\,\sigma\sigma^\prime}(k,k^\prime,q) = - G_k G_{k+q}\, \mathcal{F}_{\sigma\sigma^\prime}^{k, k^\prime, q}\,G_{k^\prime}G_{k^\prime + q}.
\end{equation}

Within the D$\Gamma$A, local approximations\cite{Toschi2007,Toschi2011} are made for $\mathcal{F}$.  
We recall, that  the latter can be viewed as the sum of all 1PI two-particle diagrams, but the terms of this sum can be further classified  as two-particle irreducible (2PI) or reducible contributions, as shown by the following relation:
\begin{equation}\label{eq:parquet}
\mathcal{F}= \Lambda+\Phi_{pp}+\Phi_{ph}+\Phi_{\overline{ph}},
\end{equation}
which is one of the parquet equations \cite{bickers2004self}. Here,  $\Lambda$ is the (fully) 2PI vertex, while $\Phi_r$ with $r = pp,\,ph,\,\overline{ph}$ are the contributions corresponding  to  two-particle scattering processes reducible   in the particle-particle, particle-hole, transverse particle-hole channels respectively (for the sake of conciseness, all spin/momenta/frequencies indices are omitted here).

Alternatively, if one focuses on the two-particle reducibility in a given channel ($r$), the diagrammatic sum defining $\mathcal{F}$  can be classified via the corresponding Bethe-Salpeter equation:
\begin{equation}
\mathcal{F}=  \Gamma_r + \Phi_r
\label{eq:BSE}
\end{equation}
where  $\Gamma_r = \Lambda + \sum_{r^\prime \not = r}\Phi_{r^\prime}$ represents, thus, the sum of all 2PI diagrams in the {\sl specific} channel $r$.
In this framework, the D$\Gamma$A is defined by a fully local, but not-perturbative, approximation  of the 2PI vertex functions.

More specifically, in the case of the full/parquet-D$\Gamma$A\cite{Toschi2007,Valli2015,gangli2016}, $\Lambda$ is approximated to a local quantity (which can be then computed from the solution of the auxiliary AIM of DMFT)  and the (non-local) $\Phi_r$ are calculated through the parquet equations of the lattice problem. However, the numerical solution of these equations is extremely challenging and therefore further approximations are often adopted at this stage. For instance, in the  ladder-D$\Gamma$A\cite{Toschi2007,Rohringer2016} the assumption of locality is extended to the $\Gamma_r$ for every channel. This amounts to a decoupling between the three different channels, whose two-particle reducible diagrams, i.e. the $\Phi_r$, are computed through Bethe-Salpeter equations (Eq.\ref{eq:BSE}), where the $\Gamma_r \sim \Gamma^{loc}_r$ is extracted from the associated AIM and the ladder resummation is performed using the (momentum dependent) DMFT Green's function 
$G_k = \left[i\nu -\xi_{\bk}-\Sigma(\nu)\right]^{-1}$. 
We emphasize that in the case of ladder-D$\Gamma$A,  the non local contributions to $\mathcal{F}$ are computed through ladder resummations in selected channels.
It is useful, thus, to introduce the auxiliary quantities $F_r \equiv \Gamma^{loc}_r +\Phi_r$,  which we will refer to, generically, as ``ladders", in the corresponding channel. 
\begin{table}
\large
\begin{tabular}{| m{0.3\columnwidth} | m{0.3\columnwidth} || m{0.3\columnwidth} |}
\hline
&&\\
& \centering $U < 0$ & {\ \ \  $U > 0$} \\
\hhline{|=|==|} 
\multirow{2}{*}{\hspace{0.55cm} local} & \multicolumn{2}{c|}{$\Lambda$,\, $\Gamma_r$ }\\
\cline{2-3}
& \hspace{0.75cm} $\Phi_{\overline{ph}}$ &  \hspace{0.75cm}$\Phi_{pp}$ \\
\hhline{|=|==|} 
\multirow{2}{*}{\hspace{0.15cm} non-local} & \multicolumn{2}{c|}{$\mathcal{F}$,\, $\Phi_{ph}$}\\
\cline{2-3}
& \hspace{0.75cm} $\Phi_{pp}$ & \hspace{0.75cm} $\Phi_{\overline{ph}}$ \\
\hline
\end{tabular}
\caption{Schematic visualization of the different approximations at the level of the two-particle vertices made within the ladder-D$\Gamma$A for the attractive ($U < 0$) and the repulsive ($U > 0$) cases. The subscript of $\Gamma_r$ refers to all three scattering channels $r = ph,\,\overline{ph},\, pp$.}
\label{tab:approx}
\end{table}
In practice, for $U>0$, one chooses to introduce the corresponding  non-local corrections in all channels of the particle-hole sector, i.e., magnetic and charge. This choice includes, in fact, the predominant magnetic fluctuations\cite{Gunnarsson2015}, while the inclusion of the charge channel ensure the validity of specific crossing symmetric relations.
Such a choice corresponds, evidently, to keep the first two terms on the r.h.s. of Eq.~\ref{eq:parquet} as fully local, which leads after some analytic manipulations exploiting  Eq.~\ref{eq:BSE},  to the following approximate expression for $\mathcal{F}_{\uparrow\downarrow}$:
\begin{equation}\label{Fupdo:Attr}
\mathcal{F}_{\uparrow\downarrow}\sim  \frac{1}{2}\left[F_d -3F_m -2\mathcal{F}^{loc}_{\uparrow\downarrow}\right]
\end{equation}
where $F_{d/m} = F_{ph\,\uparrow\uparrow}\pm F_{ph\,\uparrow\downarrow}$ and
$\mathcal{F}^{loc}_{\uparrow\downarrow}$ is the 1PI vertex function calculated from the AIM 

In an analogous way, by applying the same procedure to the attractive case ($U<0$), one must choose the channels for which the non-local  ladder treatment beyond DMFT is required. Clearly, in this case, the particle-particle sector needs to be treated in D$\Gamma$A, because it describes pairing fluctuations  which are certainly relevant for $U <0$. This consideration also applies to particle-hole charge fluctuations, so that the natural ladder approximation of Eq.~\ref{eq:parquet} will read
\begin{equation}\label{Full:Vertex:appr}
\mathcal{F}\sim \Lambda^{loc}+ \Phi_{pp}+\Phi_{ph}+\Phi^{loc}_{\overline{ph}}, 
\end{equation}
whereas, in addition to the fully 2PI vertex,  also the particle-hole transverse channel is treated at the local (DMFT) level, since it explicitly describes the (plausibly suppressed) magnetic fluctuations. Then, using the relation  $\Lambda^{loc} +\Phi_{\overline{ph}}^{loc} +\Phi_{pp}^{loc}=\Gamma_{ph}^{loc}$, we can rewrite Eq.(\ref{Full:Vertex:appr}) in the following way:
\begin{equation}\label{Fupdo:Attr}
\mathcal{F}_{\uparrow\downarrow}\sim \frac{1}{2}\left[2(F_{pp\,\uparrow\downarrow}-\mathcal{F}^{loc}_{pp\,\uparrow\downarrow}) +F_{d}-F_m\right],
\end{equation}
where $\mathcal{F}^{loc}_{pp\,\uparrow\downarrow}$ is the 1PI vertex function calculated from the AIM in the $pp$ notation. This last quantity is exact and contains the diagrams belonging to every channels at the local level, and in this case the subscript $pp$ refers only to a frequency notation, i.e. $\mathcal{F}^{loc}(\nu,\nu^\prime,\omega)=\mathcal{F}^{loc}_{pp}(\nu,\nu^\prime,\omega+\nu+\nu^\prime)$. 

We summarize the different approximations made in the ladder D$\Gamma$A for the repulsive and the attractive interactions in Table \ref{tab:approx}. While the choice of these approximations are dictated by our physical intuition, there are relevant subtleties to be considered to preserve the mapping of the system under the transformation $U \leftrightarrow -U$ at half-filling. The problem to be faced is that the fundamental building blocks of the (D$\Gamma$A) ladder resummation, i.e. the 2PI vertices $\Gamma_r$ (Eq.\ref{eq:BSE}), do \emph{not} map onto each other when $U \leftrightarrow -U$. Hence, the mapping properties of a given ladder approximation for the half-filled system must be proven explicity in this context (see Sec.\ref{sec:heuristic}).

\begin{figure}
\includegraphics[width = \columnwidth]{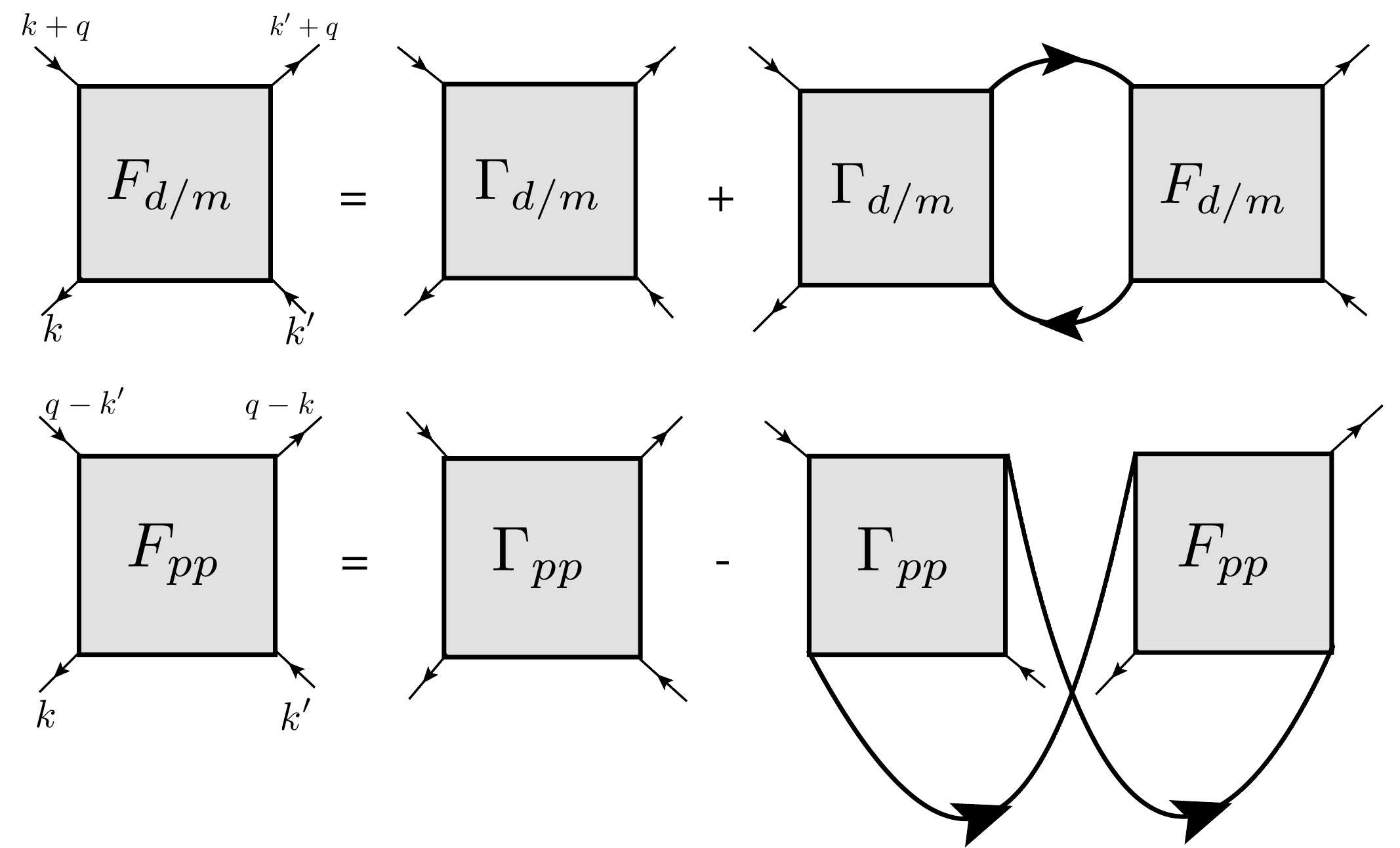}
\caption{Bethe-Salpeter equations for the particle-hole and particle-particle channels.}
\label{fig:bs}
\end{figure}

As a last step, our ladder D$\Gamma$A expressions for ${\mathcal F}$ will be inserted in the Schwinger-Dyson Eq.(\ref{eq:SDE}), to obtain the corresponding self-energy, which -for the two different cases- read:
\begin{widetext}
\begin{eqnarray}
\label{Sigma:rep}
\Sigma(k) &=& \frac{|U|}{2}n  -\frac{|U|}{2\, {\beta^2V}}\sum_{\nu^\prime q}G_{k+q}\,\chi^{\nu^\prime\, q}_0\left[-F_d^{k\, k^\prime q}+3F_m^{k\, k^\prime q}+2\mathcal{F}^{\nu\nu^\prime \omega}_{\uparrow\downarrow}\right],\  U > 0,\\
\label{Sigma:attr}
\Sigma(k) &=& -\frac{|U|}{2}n  - \frac{|U|}{2\, {\beta^2 V}}\sum_{\nu^\prime q}G_{q-k}\,\chi^{\nu^\prime\, q}_{0 \,pp}\,2\left[F_{pp}^{k \,k^\prime q}-\mathcal{F}_{pp}^{\nu\, \nu^\prime \omega}\right] -  \frac{|U|}{2\, {\beta^2 V}}\sum_{\nu^\prime q}G_{k+q}\,\chi^{\nu^\prime\, q}_0\left[F_d^{k\, k^\prime q}-F_m^{k\, k^\prime q}\right],\ U < 0 
\end{eqnarray}
\end{widetext}
where $\mathcal{F}^{\nu\nu^\prime \omega}_{\uparrow\downarrow} \equiv\mathcal{F}_{\uparrow\downarrow}^{loc\,\nu\nu^\prime \omega} $, $\mathcal{F}^{\nu\nu^\prime \omega}_{pp} \equiv\mathcal{F}_{pp\,\uparrow\downarrow}^{loc\,\nu\nu^\prime \omega} $, $F^{k\,k^\prime q}_{pp} \equiv F_{pp\,\uparrow\downarrow}^{k\,k^\prime q} $, $\chi_0^{\nu\ q}\equiv -\frac{1}{V}\sum_\bk G_{k}G_{k+q}$,  $\chi_{0\,pp}^{\nu\ q}\equiv -\frac{1}{V}\sum_\bk G_{k}G_{q-k}$. We notice that the ladders appearing in Eqs.(\ref{Sigma:rep},\ref{Sigma:attr}) depend only on the exchanged momentum $\bq$ because of the assumption of locality of $\Gamma_r$ and of the form of the Bethe-Salpeter equations, shown schematically in Fig.(\ref{fig:bs}). Moreover, the Green's function is calculated using DMFT. Nevertheless, we shall continue to use the four-component notation, in order to make a better connection between ladders and exact quantities, as will be clear in the next subsection. The first equation obviously corresponds to the typical expression exploited in previous ladder D$\Gamma$A works, while the second expression, derived here, should represent its counterpart to be used for $U <0 $ (the additional inclusion of Moriya corrections\cite{Katanin2009,Rohringer2016} will be addressed in the following,  see Sec.(\ref{sec:phys})).

The correspondence between the attractive and repulsive models is not immediately visible in the final expressions for the D$\Gamma$A self-energies  Eqs.~(\ref{Sigma:rep},\ref{Sigma:attr}), and, thus,  it has to be demonstrated explicitly.
In the next subsection we shall present a heuristic argument based on the structure of Eqs.~(\ref{Sigma:rep},\ref{Sigma:attr}), while the analytical demonstration is presented in Appendix \ref{Proof}.

\subsection{Attractive-Repulsive Mapping for ladder approximations}\label{sec:heuristic}

We consider the partial particle-hole transformation\cite{ShibaT} acting only on down fermions:
\begin{equation}
\label{eq:shiba}
\cc_{\bR\downarrow}\to (-1)^R\cd_{\bR\downarrow},
\end{equation}
which yields a perfect mapping between the repulsive and the attractive Hubbard model ($U\to -U$) at half-filling.
Specifically, this transformation maps the spin into the the pseudo-spin operators, and viceversa:
\begin{eqnarray}
U &\longleftrightarrow&-U \nonumber \\
(-1)^R\hat{S}^x(\bR)&\longleftrightarrow&\hat{S}_p^x(\bR)\nonumber \\
(-1)^R\hat{S}^y(\bR)&\longleftrightarrow&\hat{S}_p^y(\bR)\nonumber \\
(-1)^R\hat{S}^z(\bR)&\longleftrightarrow&(-1)^R\hat{S}_p^z(\bR)
\end{eqnarray}
where, $\hat{S}^{\alpha}(\bR) \equiv \psid_{\bR}\,\tau^{(\alpha)}\,\psi^{\ }_{\bR} $ and $\hat{S}^{\alpha}_{p}(\bR)\equiv \phid_{\bR}\,\tau^{(\alpha)}\,\phi^{\ }_{\bR}$ are respectively the spin and pseudo-spin operators, with $\psi_\bR \equiv (\cc_{\bR\uparrow},\cc_{\bR\downarrow})$, $\phi_\bR \equiv (\cc_{\bR\uparrow},\cd_{\bR\downarrow})$, and $\tau^{(\alpha)}$ are the Pauli matrices with $\alpha = x,y,z$.   As a consequence, the components of the staggered magnetization along the $xy$-plane are mapped onto the superconducting (SC) order parameter, that is two dimensional, and the staggered magnetization along the $z$-axis is mapped onto the Charge Density Wave (CDW) order parameter. 

 The exact mapping  ($U\to -U$) of the half-filled case implies as direct consequence --particularly relevant for our purposes--  that  $\Sigma(k)$ does not depend on the sign of the interactions.
At the same time, the validity of such a basic requirement should be preserved by any approximation aiming to provide a {\sl consistent} description of the repulsive and the attractive sector. This should obviously apply also to the case of the ladder D$\Gamma$A.

For the exact case, briefly discussed in Appendix \ref{Proof}, the mapping of the self-energy, expressed via the Schwinger-Dyson equation of motion, is directly guaranteed by the following property of the exact-1PI vertex function:
\begin{equation}
\label{eq:mapF}
\mathcal{F}_{(U),\uparrow\downarrow}^{k,k^\prime,q}  = -\mathcal{F}_{(-U),\uparrow\downarrow}^{k,-q-k^\prime-\Pi,q}, 
\end{equation}
where $\Pi = (\boldsymbol{\Pi},0)$, with $\boldsymbol{\Pi}\equiv(\underbrace{\pi,\pi,...,\pi}_d)$.
On the other hand, in our ladder D$\Gamma$A approximation, we express $\mathcal{F_{\uparrow\downarrow}}$ as a function of non-local ladders relative to different scattering channels in an independent way (for $U> 0$ and $U <0$). 




We observe, that in both the RHSs of Eqs.(\ref{Sigma:rep},\ref{Sigma:attr}),  {\sl four} convolutions involving a momentum-dependent ladder $F_r$, appear. This corresponds to the inclusion of non-local fluctuations in the spin and pseudospin sectors,  and can be regarded, loosely speaking, as the \emph{non-local degrees of freedom} treated by the  ladder schemes. 

In fact, in the repulsive case, the non-local ladders are built  for the {\sl three} spin components of magnetic channel as well as for {\sl  one} component of the pseudospin  sector (i.e.,  in the charge sector). For $U <0$, instead, the situation is perfectly mirrored: the ladder are built for the {\sl three} components of the pseudospin fluctuations (i.e., in the $pp$ and in the charge sector), and for {\sl one} of spin component. As this is consistent with the mapping of the spin/pseudospin operators through the Shiba transformation, it  should also guarantees, in principle, the perfect equivalence between the corresponding ladder approximations for $U > 0$ and and $U < 0$.

One should consider, nonetheless,  that the $F_r$ appearing in the approximated expressions for $\Sigma(k)$ are complicated functions of the frequencies and momenta, and, that, in general, it is not true that $F^{k\,k^\prime q}_m{\to}F_{d}^{k\,k^\prime q}$ under the transformation $U\to-U$. The explicit proof of the equivalence of the ladder approximations (with mirrored non-local fluctuating channels) is provided in Appendix \ref{Proof}, whose main result is recalled below. This is given by the following relation fulfilled by ladders:
\begin{equation}\label{ladder:updo:mapping}
F^{k,k^\prime,q}_{(U),\,\uparrow\downarrow} = -F^{k,-q-k^\prime-\Pi,q}_{(-U),\, \uparrow\downarrow},
\end{equation} 
 that is the same of Eq.(\ref{eq:mapF}). This guarantees that ladders map as the exact full vertex under the partial particle hole transformation defined in Eq.(\ref{eq:shiba}). Thus,   since the particle-hole symmetry requirements  are fulfilled at the two-particle  level (accordingly to Eq.({\ref{ladder:updo:mapping}})) these properties remained preserved through the Schwinger-Dyson Equation of motion also for the self energy.
As similar considerations will hold for a generic ladder based approximation, our derivation might be useful to develop consistent extensions of other ladder-based methods (e.g. DF, DMFT+ FLEX, 1PI) for treating systems with contact  attractive interactions.

The fulfilment of the Shiba mapping  has been also numerically verified by comparing the D$\Gamma$A self-energy at half-filling for the half-filled attractive and repulsive models.

In both cases, the DMFT vertex and self-energy, computed with an Exact Diagonalization solver (with 5 sites), have been used for computing the input quantities of the ladder D$\Gamma$A.
In all the calculations we used a frequency grid for local vertices of 40-60 Matsubara frequencies for the subsequent D$\Gamma$A calculation,  and a momentum grid for the internal bubble of 40-80 momenta.




\subsection{SU(2) spin/pseudospin symmetries in the ladder-D$\Gamma$A}\label{sec:symmetries}

After demonstrating the validity of the mapping at the level of our ladder D$\Gamma$A scheme, a comment is due about the treatment  of the internal symmetries of the problem, with particular reference, here, to the SU(2) symmetry of the spin and the pseudospin variables. This represents, in fact, a somewhat subtle issue to be considered for approximations based on the selection of channels (as we do in ladder D$\Gamma$A and/or DF), which, however, has been never discussed explicitly in the context of the diagrammatic extensions of DMFT. Our comparison between ladder approximations for $U> 0$ and $U<0$ represents, thus, an ideal playground to fill this gap. 

As it is known, the particle-hole symmetric single-band Hubbard hamiltonian displays a SU(2)-symmetry  both for the spin and the pseudospin variables. Thus, it is quite evident that the choice of considering non-local ladders for all (four) channels of the particle-hole sectors, made in standard ladder D$\Gamma$A approximations for the repulsive case, corresponds to a violation of the SU(2) symmetry for the pseudospin variables: This is due to  the different  (nonlocal vs. local) treatment of the charge- and the particle-particle-channel, although they would correspond to "equivalent" fluctuations of the pseudospin variables. 

For analogous reasons, we observe a violation of the SU(2) spin symmetry in the attractive case. In fact,  our ladder D$\Gamma$A approximation for the AHM has been designed to be physically equivalent to the corresponding one for $U>0$. 
This violation appears in the selection of a {\sl single} component for which the spin fluctuations are considered nonlocal. 

It is important to emphasize that the effect of such violations on the final results of a ladder D$\Gamma$A calculations are typically marginal or completely negligible, because they always affect by construction the secondary/suppressed fluctuating channels.  In other words, if the violation is large, this means that the corresponding ladder approximation is not suited for the physical problem under consideration.

Nonetheless, one might wonder how to completely circumvent this problem in future, further improved ladder calculation schemes.
Hence, in order to avoid any kind of artificial asymmetry, the non-local treatment should be extended to {\sl all} the $6$ ladders appearing in the expression of scattering amplitude ${\cal F}_{\uparrow\downarrow}$, i..e. in Eq.(\ref{Full:Vertex:appr}).

This would be possible at the level of a fully nonlocal ladder treatment, where, however, one should continue to neglect the mutual influence between the six non-local channels, in order to still avoid the complexity of parquet-based algorithms.

 \section{Numerical results: The 3D model out of half-filling}\label{sec:III}
 
 In this section, we apply the ladder D$\Gamma$A Eq.(\ref{Sigma:attr})  to study the three dimensional AHM in the more general non particle-hole symmetric situation ($\mu \ne \frac{U}{2}$), considering the case of densities $n$ lower than half-filling. In particular, we focus on the SC and CDW instabilities.
 We recall that, at half-filling CDW and SC phases are fully degenerate and they obviously share the same critical temperature and the critical exponents of the $3d$ Heisenberg model universality class. 
  When the system is doped, CDW is progressively suppressed and SC state emerges as the only stable broken symmetry phase\cite{Micnas1990}.  
  
 The critical temperatures related to the different instabilities are obtained by calculating the associated physical susceptibility and tracking their divergence as a function of temperature. We also note that, in ladder D$\Gamma$A,  the different channels (see Eqs.(\ref{BS:ph},\ref{BS:pp})) are not coupled and, thus, one can study their associated susceptibilities independently.  This gives us the possibility of tracking, separately,  the divergences of the susceptibility in both  the leading particle-particle and the sub-leading charge channel,  and to determine, thus, the  two correspondent critical lines in the phase diagram $T$ vs $n$ at fixed $U$.
As mentioned, we expect $T_c^{SC}\geq T_c^{CDW}$ for every filling, though the situation could be reversed if retarded interactions or a mass imbalance between the two fermionic species (here: $\uparrow$ and $\downarrow$ electrons)  had been considered. Hence, though it refers here to a sub-leading instability,  our numerical evaluation of $T_c^{CDW}(n)$ could represent a useful reference point in view of Fermi mixtures of ultra cold atoms.

\subsection{Evaluation of physical susceptibilities}\label{sec:phys}

The physical susceptibilities, depending on {\sl a single} momentum index, can be directly derived  from  the so-called generalized susceptibilities, depending on {\sl three} momenta and defined as:
\begin{equation}\label{gen:susc}
\chi_{\sigma\sigma^\prime}^{k,k^\prime,q}\equiv 
 -\mathcal{V}\delta_{k,k^\prime}\delta_{\sigma,\sigma^\prime}G_{k}G_{k+q}-G_kG_{k+q}\mathcal{F}^{k,k^\prime,q}_{\sigma,\sigma^\prime}G_{k^\prime}G_{k^\prime + q},
\end{equation}
by means of a summation over the fermionic indices
\begin{equation}\label{phys:susc}
\chi_{\sigma\sigma^\prime}(q) = \frac{1}{\mathcal{V}^2}\sum_{k,k^\prime}\chi_{\sigma,\sigma^\prime}^{k,k^\prime,q}.
\end{equation}

 
 The calculation of  $\chi^{DMFT}_{\sigma,\sigma^\prime}(q)$ in finite dimensions represents the first step in the ladder D$\Gamma$A algorithm. As DMFT is an exact theory in $d= \infty$ only, its two (and many-)particle self-consistency with the local properties of the auxiliary AIM is violated in finite dimension. For instance the equivalence $\frac{1}{V}\sum_{\bq}\chi^{DMFT}_{\sigma,\sigma^\prime}(q) = \chi^{loc}_{\sigma,\sigma^\prime}(\omega)$, where $\chi^{loc}$ of the auxiliary impurity susceptibility is  {\sl no} longer guaranteed by the validity of the corresponding one-particle self-consistence condition on the Green's function ($\frac{1}{V}\sum_{\bk}G^{DMFT}_k = G^{loc}(\omega)$).
 As a consequence the physical susceptibility violates certain sum rules holding in the exact case, and the self-energy in Eqs.(\ref{Sigma:rep},\ref{Sigma:attr}) acquires an incorrect asymptotic behavior at high-frequencies\cite{Katanin2009,rohringer2013new, Rohringer2016}. 
 
At the level of the ladder D$\Gamma$A this problem can be solved\cite{Katanin2009,Rohringer2016} by including a variational parameter $\lambda_{\sigma,\sigma^\prime}> 0$ in the definition of the physical susceptibility: 
\begin{equation}\label{phys:susc:DGA}
 \chi^{D\Gamma A}_{\sigma,\sigma^\prime}(q) \equiv\left[ \left(\chi^{DMFT}_{\sigma,\sigma^\prime}(q) \right)^{-1} + \lambda_{\sigma,\sigma^\prime}\right]^{-1}.
\end{equation}
This $\lambda_{\sigma,\sigma^\prime}$  correction term is chosen\cite{Katanin2009} in order to restore the two-particle self-consistency of the theory: 
\begin{equation}\label{susc:cond}
\frac{1}{\mathcal{V}}\sum_{q}\chi^{D\Gamma A}_{\sigma,\sigma^\prime}(q)\overset{!}{=}T\sum_{\omega}\chi^{loc}_{\sigma,\sigma^\prime}(\omega),
\end{equation}
and thus the correct high-energy asymptotics of $\Sigma$\cite{Katanin2009,rohringer2013new}, though other choices\cite{Rohringer2016} are possible. 
 
 The role of the variational parameter $\lambda$ can be most easily understood by considering a system is close to a phase transition:
If $\chi^{DMFT}$  has a divergence at the point $Q = (\mathbf{Q},0)$ at the critical temperature $T^{DMFT}_c$, for $T \sim T^{DMFT}_c$ and  $q\sim Q$, one can write the susceptibility as $\chi^{D\Gamma A}(q)\sim A\left[(q-Q)^2 + (\xi^{-2}+\lambda) \right]^{-1}$, where $\xi$ is the coherence length and $A$ a non universal constant. Hence, a parameter $\lambda > 0$ lifts the divergency of the physical susceptibility, reducing the corresponding critical temperature.  Similarly as in TPSC \cite{vilk1997non}, which is based on enforcing the same self-consistent condition in a perturbative (RPA-like) framework, in $d=2$ one recovers the vanishing critical temperature predicted by the Mermin-Wagner theorem and expects significant corrections in $d=3$. The specific results for the AHM are presented below. 

We stress that given the $SU(2)$-symmetry of the problem, the $\lambda$-corrections, Eqs.(\ref{phys:susc:DGA},\ref{susc:cond}), evaluated within different channels are independent on each other. Hence, the critical temperature is reduced only by  the (non-local) fluctuations of the order parameter.

 \begin{figure*}[htbp!]
\includegraphics[width = \textwidth]{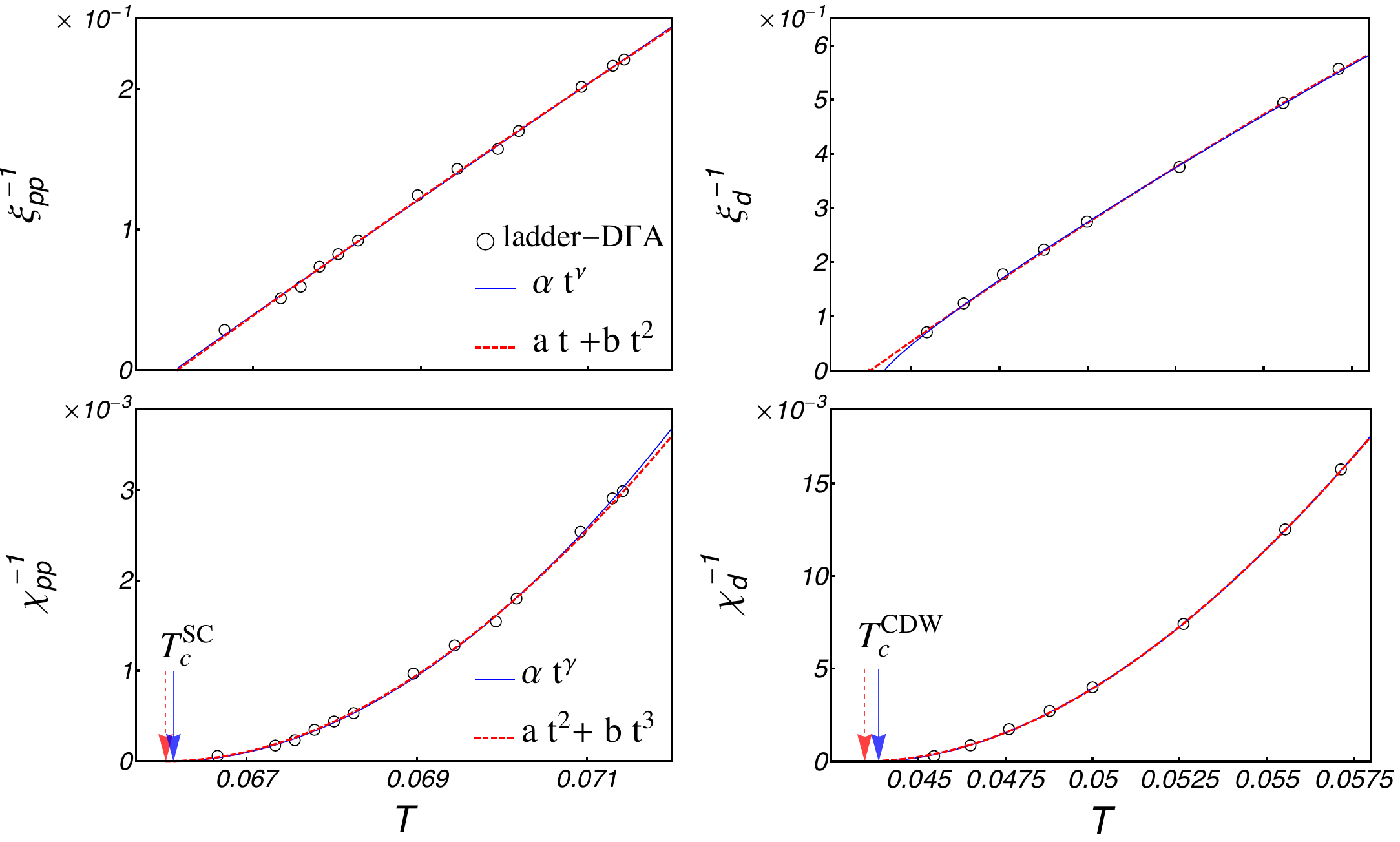} 
\caption{Numerical calculation of the particle-particle (charge) susceptibility $\chi_{pp}(0)$ ($\chi_d(\Pi)$) and correlation length $\xi_{pp}$ ($\xi_d$) computed in ladder-D$\Gamma$A as a function of the temperature for $n = 0.5$. In order to extrapolate the critical temperature of the SC (CDW) instability we performed a numerical fit where the critical exponent is taken as a free parameter (blue solid line). We also performed a fit with the exponent of the $O(\infty)$ universality class (red-dashed line) of our data taking into account of sub-to-leading orders (see the main text). 
A review of our results compared with the critical exponents of the different universality classes is shown in Table \ref{tab:exps}.
 The vertical arrows mark the corresponding estimates of the critical temperatures, as obtained by means of the two fitting procedures.}
\label{fig:fit}
\end{figure*}

\begin{table*}
\begin{tabular}{|m{0.04\textwidth}|| m{0.11\textwidth}|m{0.1\textwidth} | m{0.1\textwidth} |m{0.11\textwidth} |m{0.11\textwidth} ||m{0.11\textwidth}| m{0.11\textwidth}||m{0.11\textwidth}|}
\hline
 & $\hspace{0.5cm} d = \infty$&\centering $O(1)$ &\centering$O(2)$&\centering$O(3)$&\centering $O(\infty)$&\centering ladder-D$\Gamma$A&  \centering ladder-D$\Gamma$A& \ ladder-D$\Gamma$A\\
 &\centering (Mean-Field)&\centering (Ising)& \centering(XY)& \centering(Heisenberg)&\centering (TPSC) & \centering SC&\centering CDW & \hspace{0.5 cm} AFM\cite{Rohringer2011}
 \\
\hline
&&&&&&&&\\
  $\hspace{0.25cm } \nu$&\centering 1/2&\centering 0.63&\centering 0.67 & \centering 0.71&\centering 1 & \centering 0.98&\centering 0.86& \hspace{0.6 cm} 0.72 \\
\hline
&&&&&&&&\\
   $\hspace{0.25cm }\gamma$&\centering 1&\centering 1.23&\centering 1.33 & \centering 1.38&\centering 2 & \centering 1.9&\centering 1.79&\hspace{0.6cm} 1.41 \\
  \hline
\end{tabular}
\caption{Comparison between the critical exponents belonging to different universality classes and the ones obtained in ladder-D$\Gamma$A. We observe that our estimates, extracted from the single power-law fits shown in Fig.(\ref{fig:fit}), relative to the SC and CDW instabilities are closer to the \cite{vilk1997non,am-dare1996} results than to the expected values of the critical exponents of respectively the XY and Ising universality classes   (see text).}
\label{tab:exps}
\end{table*}

\ \\
\subsection{Critical properties and ordering temperatures}\label{sec:critical}

The determination of the ordering temperature ($T_c$) for the different second-order phase transitions in D$\Gamma$A implies a 
simultaneous definition of the associated critical properties. This is necessary, because $T_c$ is computed, similarly as in DMFT,
 through the divergence of the corresponding static susceptibility $\chi$ and the correlation length $\xi$. However, differently from the DMFT case, where their critical behavior [$\chi(T) = (T- T_c)^{-\gamma}$, $\xi = (T- T_c)^{-\nu}$] is known {\sl a priori}, being purely mean-field ($\gamma=1, \nu =\frac{1}{2}$), the non-local spatial correlations captured by the D$\Gamma$A modify it considerably, yielding\cite{Rohringer2011} larger values of both critical exponents in $d=3$. In practice, the determination of $T_c$ in D$\Gamma$A is performed by fitting the numerical data of $\chi$ in the proximity of the transition, i.e., where the value of $\chi$ and $\xi$ are large enough to allow to enter the critical region (but still in the range numerically treatable in the momentum grid of the calculation).

In particular, when $T\sim T_c$ and $q \sim Q$, the susceptibility  will be expressed as :
\begin{equation}\label{coherence:l}
\chi^{D\Gamma A}_{d/pp}(q) \sim \frac{A_{d/pp}}{(q - Q)^2+\xi_{d/pp}^{-2}},
\end{equation} 
where $Q= (\mathbf{0},0)$ for the particle-particle channel, while for the charge sector $Q = (\boldsymbol{\Pi} - \boldsymbol{\delta}(n),0)$, with $\boldsymbol{\delta}(n)$ being different from zero in a very narrow region of filling values where incommensurate order occurs \cite{sergioC1993}, and we have a splitting of the susceptibility peak into  multiple peaks, 
accordingly to the multiplicity of $\boldsymbol{\Pi} - \boldsymbol{\delta}(n)$.
 Eq.(\ref{coherence:l}) can be exploited, thus,  to compute the correlation length via a fit of the numerical data in momentum space. 
 This way, the temperature-dependence of both $\chi$ and $\xi$ is determined, and, subsequently used to evaluate the  critical  exponents and the critical temperature within D$\Gamma$A. 
 
 We recall, in this respect, that previous D$\Gamma$A studies\cite{Rohringer2011,rohringer2018diagrammatic} for the repulsive Hubbard model in $3d$ have yielded a plausible reduction of the antiferromagnetic ordering temperature ($T_N$) and values of the critical exponents ($\gamma \sim  1.4$, $\nu \sim 0.7$, see Table \ref{tab:exps}) arguably consistent with the $3d$-Heisenberg universality class\cite{zinn1996quantum} (the expected one). Similar results have been obtained\cite{Antipov2014,Hirschmeier2015} with other diagrammatic extensions of DMFT, as  the Dual Fermion, both for the CDW transition of the Falicov-Kimball in different dimensions, in good agreement with Ising universality class, and, again,  for the AF  transition of the $3d$ Hubbard model, giving exponents close to the Heisenberg model ones. 
 
Therefore we would expect that the D$\Gamma$A applied to the AHM would also yield critical exponents numerically consistent with the exact ones for the corresponding universality classes.  As we are considering here the case of a not half-filled (i.e.,  non particle-hole symmetric) AHM, the SC  and the CDW transition are no longer degenerate:  the expected universality classes are, thus
 $3d$-XY  ($\gamma \simeq 1.32 $; $\nu \simeq  0.67 $) for the SC transition and $3d$-Ising  ($\gamma \simeq 1.23 $; $\nu \simeq  0.63$) for the CDW \cite{zinn1996quantum}.  
 
 \begin{figure*}[th!]
\includegraphics[width = \textwidth]{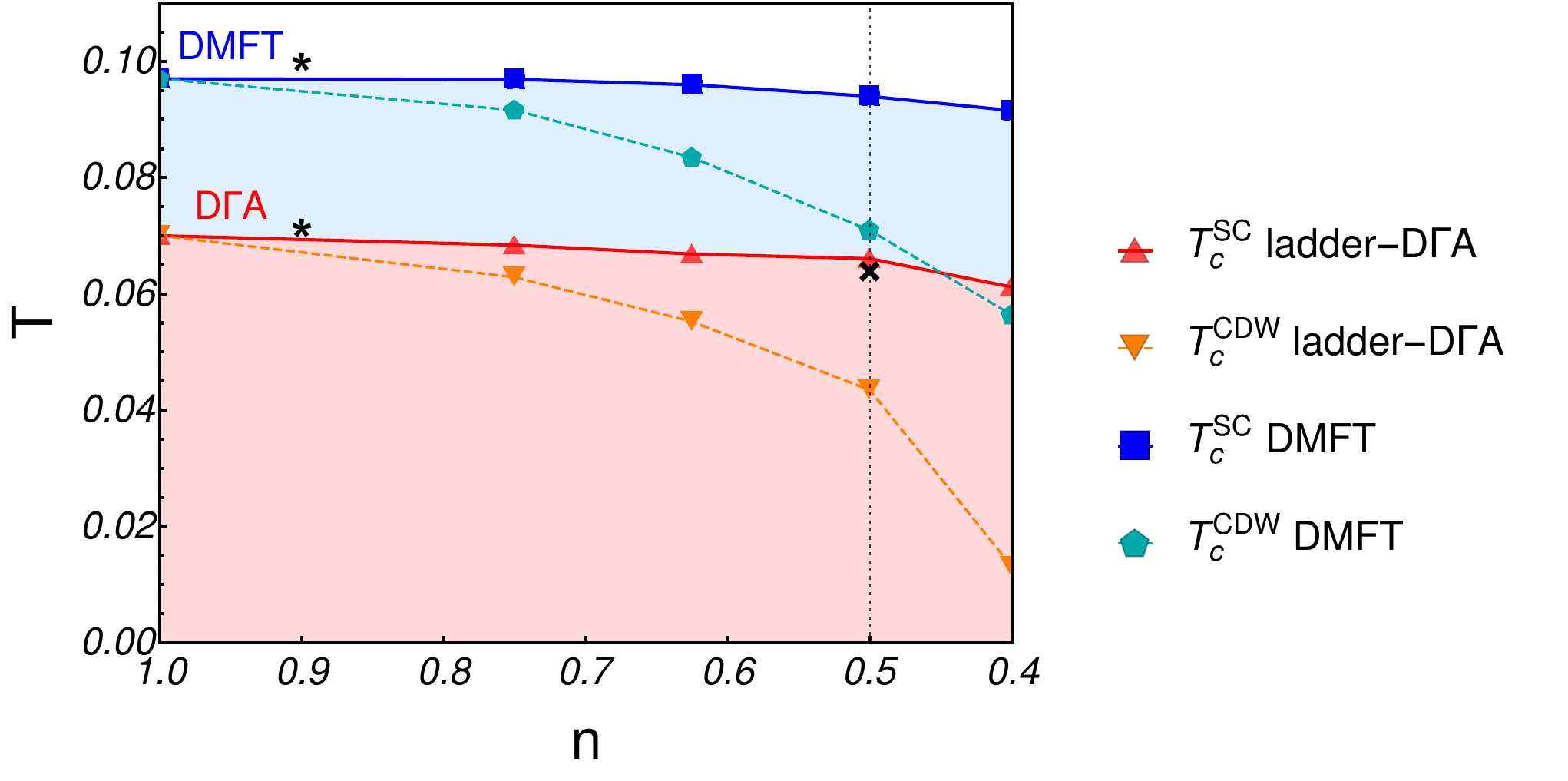}
\caption{Phase diagram in the plane ($n$,$T$) for the fixed value of the onsite interaction $U = -2.0$. The solid lines refer to the critical temperature relative to the SC ordering estimated in DMFT (blue squares) and in D$\Gamma$A (red triangles). The dashed lines refer to the critical temperature relative to the CDW instability evaluated in DMFT (cyan diamonds) and in D$\Gamma$A (orange triangles). The vertical thin line refers to $n=0.5$, that is the filling relative to the data shown in Fig.(\ref{fig:fit}). Asterisks refer to the points where the self-energy shown in Fig.(\ref{fig:selfdga}) has been calculated. 
The cross lying on the vertical line at $n = 0.5$ marks the critical temperature of superconductivty evaluated using lattice    QMC\cite{Sewer2002,Burovski2006}.} 
\label{PhD}
\end{figure*}

 As we will see below, however, our D$\Gamma$A results for the AHM, both for the SC and the CDW transition,  appear not consistent with the corresponding universality classes. 
More specifically, in Fig.(\ref{fig:fit}), we show the numerical data for correlation lengths and susceptibilities of the two different channels under scrutiny.  If assuming a simple power-law form, as done in Refs.~\cite{Rohringer2011,Antipov2014,Hirschmeier2015},  the fit yields exponents systematically larger than the ones belonging to the expected universality classes. This is particularly evident for the case of the CDW, where the difference between  fitted values ($\nu_{d} \sim 0.86$, $\gamma_{d}\sim1.79$) and the smaller exponents of the $3d$-Ising universality class is even more evident (see Table \ref{tab:exps}).
These systematically higher values might suggest, instead, a consistency with the critical exponents  of the spherically symmetric (Kac\cite{stanley1971phase}) model. These are $\nu=1$, $\gamma=2$ (with vanishing anomalous exponent $\eta =0$), which also coincides with the exponents found in TPSC \cite{vilk1997non,am-dare1996}. In this scenario, the residual difference between the fitted exponents and those belonging to the $O(\infty)$ universality class might be ascribable to the effects of sizable next-to-leading terms in the expansion around $t= (T- T_c)/T_c$.  For instance, by considering the   sub-to-leading orders\cite{PhysRevB.85.201101}, one would   $\xi^{-1} = a_{\xi}\,t + b_{\xi}\,t^2$   and $\chi^{-1} = a_{\chi}\,t^2 + b_{\chi}\,t^3$, which allows for a good matching fit of our D$\Gamma$A data for $T \sim T_c$.

In any case, our results show that the different fitting procedures associated with different critical exponents yield no significant variation of the estimated critical temperatures (cf. vertical arrows in the bottom panels of Fig.(\ref{fig:fit})). Hence, the results obtained for the phase-diagram of the AHM in ladder-D$\Gamma$A can be considered quite accurate. 
\begin{figure*}[hbtp!]
\includegraphics[width= \textwidth]{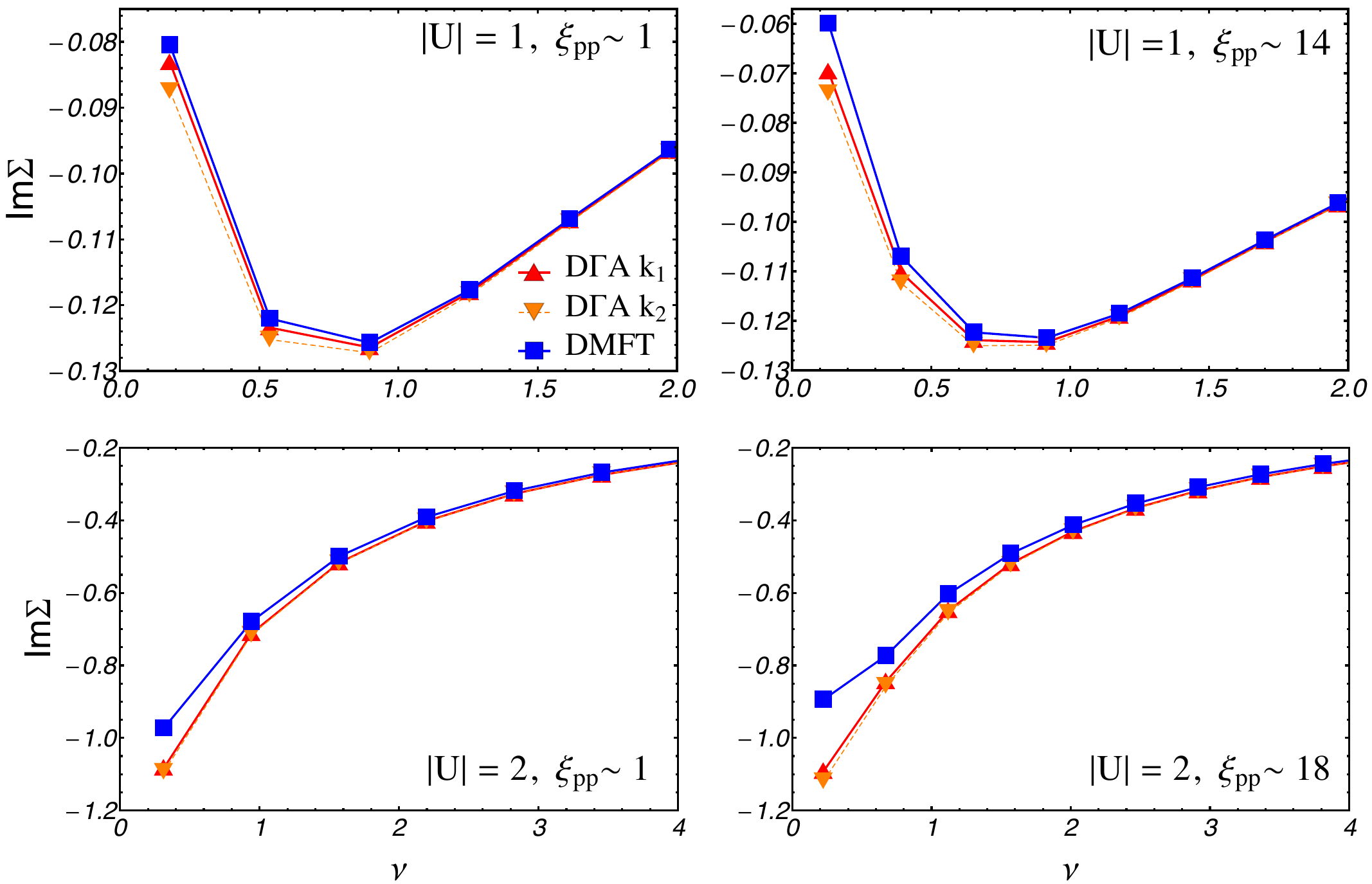}
\caption{Imaginary part of the self-energy evaluated on the imaginary axis for different values of the onsite interaction and temperature. Specifically, we consider (upper panel) $U = -1$ , $n =0.95$ and temperatures $\beta = 17.5,\ 24$ (from left to right)  and (lower panel) $U = -2$,  $n = 0.9$ and temperatures $\beta = 10,\ 14$ (from left to right).  The D$\Gamma$A self-energy has been reported for two $\bk$-points, namely $\bk_1$ and $\bk_2$, chosen among all the points lying on the Fermi Surface in the way to maximize the spread of the self-energy evaluated at the first Matsubara frequency, i.e. to maximize the following quantity $|\mbox{Im}\Sigma(\pi T,\bk_1)-\mbox{Im}\Sigma(\pi T,\bk_2)|$. The coherence length    of the pairing fluctuations $\xi_{pp}$ is explicitly reported, in order to quantify the "distance" from the SC transition of the corresponding data set: Significant deviations from the DMFT data are obtained close to the SC phase transition. The data for $U = -2$ have been calculated in points of the phase diagrams marked as asterisks in Fig.(\ref{PhD}).}
\label{fig:selfdga}
\end{figure*}

Our estimates of the critical temperatures of the SC and CDW instabilities, computed in DMFT and in ladder D$\Gamma$A for fixed interaction $U = -2.0$, and different values of the density $n$ are summarized in the  $(T,n)$-phase diagram in the plane of  Fig.(\ref{PhD}).

We observe, that critical temperature associated to the SC ordering depends very weakly from the filling, a feature, which should be indeed expected, if one consider that the doped AHM can be mapped in the half-filled repulsive one in presence of a magnetic field. 
In fact, this trend has already been shown in previous studies at the level of DMFT \cite{metznerAttractive,Toschi2005-1,agnese2016}, and it is preserved also when the spatial correlations beyond DMFT are taken into account by means of ladder D$\Gamma$A.  Nonlocal correlations, however, induced a significant reduction of the critical temperature, of about the 30$\%$ w.r.t. the DMFT one, in a similar fashion to what it is observed at half-filling\cite{Rohringer2011,Rohringer2016,Hirschmeier2015}.
 
The suppression  of the critical temperature due to nonlocal correlations is even more pronounced in the case of the CDW ordering, where we report a maximal reduction of about 76$\%$ w.r.t. DMFT for the highest doping (lower density) we considered in our study. 
A greater suppression of the ordering tendency in the charge channel is expected on a physical basis, since CDW ordering (associated in the mapped model to the AF component in the direction of applied magnetic field)  should disappear for strong enough hole doping (in the mapped model: magnetic field). According to our D$\Gamma$A data, such critical value of the filling $n=n_c$, where a QCP might appear, could be significantly reduced w.r.t. the DMFT estimate for three dimensional systems.
Finally, let us mention, that we expect the CDW to become incommensurate for a tiny range of fillings just before the critical value ($n \sim n_c$), as already shown in the DMFT literature \cite{sergioC1993}. The determination of this elusive, tiny region, where a switch to a first order transition might eventually occur\cite{PhysRevB.56.14469}, is numerically challenging in ladder D$\Gamma$A, but it might considered in future, focused studies.

Finally, we also compared our results with with  lattice Quantum Monte Carlo (QMC) data taken from the literature \cite{Sewer2002,Burovski2006}  (see black cross in Fig.(\ref{PhD})).
We observe, in the interesting (overdoped) region of $n = 0.5$, a good quantitative agreement between the critical temperatures estimated within the two approaches, which supports the quantitative accuracy of the ladder-D$\Gamma$A at least for the evaluation of the ordering temperatures.

\subsection{The fermions self-energy}\label{sec:single}
In this section, we highlight the effect of nonlocal spatial fluctuations onto the single-particle properties of the AHM. 

In particular, in Fig.(\ref{fig:selfdga}), we show the self-energy obtained in ladder D$\Gamma$A close to half-filling, evaluated for two momenta at the Fermi-Surface ($\bk_1$, $\bk_2$) and compare it with the corresponding DMFT one. We recall that, in this part of the study, we can access only the region of the phase-diagram above the SC instability of D$\Gamma$A, because the determination of the self-energy through the ladder D$\Gamma$A, differently from the  DMFT case, requires to work in the thermodynamically stable phase\cite{Toschi2007,Katanin2009}. In fact,  in order to access the single-particle properties below the SC instability, an extension of the D$\Gamma$A equations and algorithms to the  broken symmetry phases would be necessary.

For generic values of filling and temperature  away from the SC phase transition (i.e. when $\xi_{pp} \sim 1$), the data obtained respectively using DMFT and D$\Gamma$A are very similar (left panels of Fig.(\ref{fig:selfdga})), giving additional proof of the validity of the DMFT approximation for a generic case in three dimensions.
At the same time, when $n$ and $T$ are chosen close to the SC critical line, we observe a sizable deviation from the DMFT data, 
 particularly enhanced in the intermediate coupling regime ($U = -2$, right bottom panel of Fig.(\ref{fig:selfdga})): In the low-frequency sector, the imaginary part of the self-energy is larger in absolute value using D$\Gamma$A, signaling that electrons at the Fermi-surface become more correlated. This is due to increased SC (as well as CDW close enough to half-filling) fluctuations, that in DMFT were included only at the local level.  

Furthermore, the data at intermediate coupling show a very different temperature trend of the DMFT and the D$\Gamma$A self-energies. While the former is reduced in absolute value for lower temperature (indicating a reduction of the scattering rate), the latter shows a slight increase for the lowest temperature. In fact, the temperature trend in DMFT simply reflects the tendency towards a coherent low-temperature metallic phase in the crossover region of DMFT, resulting from the large entropy associated to the local pair insulator phase. Conversely, in D$\Gamma$A this effect is gradually overcome by the progressively increased scattering due to the non-local correlations\cite{Rohringer2016}.

 It is interesting to notice, on the other hand, that we do not observe a significant dispersion of the D$\Gamma$A self-energy on the Fermi surface, i.e. as a function of the crystalline momentum. Hence, according to our ladder D$\Gamma$A results,  at intermediate coupling we have the following situation: a  SC (and CDW close enough to half-filling) susceptibility strongly depending on the momentum, which, at the Fermi level, renormalizes mostly the low-frequency dependence of the self-energy through the Schwinger-Dyson equation, and determine an enhanced scattering rate along the  whole Fermi surface.

The weak momentum-dependence of the scattering rate at the Fermi surface is a consequence of the three dimensional momentum integration in the Schwinger-Dyson equation, which partly smooths the k-selective effects of nonlocal (here: SC) fluctuations, and also a result of the relatively large value of the interaction considered, which induces a certain degrees of localization already at the DMFT level. 
We note, however, that a stronger momentum dependence (mostly in the real part of the self-energy) is found when considering momenta far away (above and below) from the Fermi surface (not shown). Furthermore,  consistently with the above considerations, a relatively larger momentum dependence in the D$\Gamma$A self-energy can be observed also at the Fermi level, when approaching the SC transition for smaller values of the interaction.

\section{Conclusions}\label{sec:concl}

In this paper we have presented an extension of the equations and the algorithm for the dynamical vertex approximation in its ladder version (ladder D$\Gamma$A), designed to treat non-local spatial correlations beyond DMFT in the attractive Hubbard model.

The derivation of the corresponding equations has been dictated by the requirement of a perfect equivalence to the ladder D$\Gamma$A scheme for repulsive interactions in the case  of the half-filled, particle-hole symmetric model. 
In particular we proved that $\Sigma(k)$ calculated within ladder-D$\Gamma$A does not depend on the sign of the interactions at half-filling as implied by the Shiba transformation. This result is not trivial for two main reasons. First,  the different approximations performed for the two particle reducible objects (see Table \ref{tab:approx}).  As we clarify for the first time in this work, this has specific consequences for the $SU(2)$ symmetry of the D$\Gamma$A solution (see Sec.\ref{sec:symmetries}), suggesting further possible improvements of the method.

The second reason is related to the fact that, in general, the $\Gamma_r$ do not map onto each other when $U \leftrightarrow -U$, in contrast with the physical susceptibilities in the different channels.
Therfore, one should pay special care when a ladder approximation is performed and the proof of the mapping becomes necessary.
In particular, the non trivial relation which ensures the mapping within a ladder approximation is the following: $F^{k,k^\prime,q}_{(U),\,\uparrow\downarrow} = -F^{k,-q-k^\prime-\Pi,q}_{(-U),\, \uparrow\downarrow},$ 
as it has be
derived explicitily Appendix \ref{Proof}.

We also stress that the main steps of our derivations, as well as the considerations made on their basis, are relevant not only for the specific case of the D$\Gamma$A, but for any  ladder diagrammatic scheme (e.g., in the context of the diagrammatic extensions of DMFT\cite{rohringer2018diagrammatic}, for the ladder DF\cite{Hafermann2009} or the 1PI approach) which needs to be extended to the attractive case.

After proving, that our implementation  preserves the physical equivalence at the  ladder D$\Gamma$A level, we have applied our modified algorithm out of particle-hole symmetric sector, studying the effects of nonlocal correlations for the more relevant case of an (hole) doped AHM in three dimensions. In particular, we have focused on the physics close to the superconducting  (SC) and charge-density-wave (CDW)  instabilities, by computing the temperature behavior of the corresponding susceptibilities and coherence lengths in the critical region, both in DMFT and ladder D$\Gamma$A.  The fitted critical exponents in ladder D$\Gamma$A  resulted larger not only than the DMFT one (as expected), but also than  the exact ones belonging to the corresponding universality classes.
In fact, the fit including also subleading terms suggests that the actual critical exponents of our ladder D$\Gamma$A calculations with Moriya corrections belong to the universality class of the spherical model (see Table \ref{tab:exps}) where:
$$\nu = 1, \ \gamma = 2,$$

which would also correspond to the result obtained using the TPSC\cite{vilk1997non}.

 Nonetheless, the determination of the critical temperature turned out to be essentially unaffected by the underlying uncertainty in the exponents, allowing for a reliable computation of the phase diagram of the AHM in three dimensions.
The obtained critical temperature computed in D$\Gamma$A are visibly reduced with respect to DMFT and TPSC, which are indeed expected to overestimate the ordering temperature (cfr. \onlinecite{rohringer2018diagrammatic}) and shows a good quantitiative agreement with lattice QMC results\cite{Sewer2002,Burovski2006}. 
 
This way we could  highlight the different effects of the non-local (SC and CDW) fluctuations on the corresponding transition temperature, as well as on the single-particle properties for the parameter regime of temperature above the physical (SC) transition. In particular, close enough to the phase transition, the absolute magnitude of the imaginary part of the fermionic self-energy increases with respect to the one calculated in DMFT, indicating an increased electronic scattering rate due to the enhancement of the non-local pairing fluctuations (see Fig. \ref{fig:selfdga}). 

The equations and the corresponding ladder D$\Gamma$A algorithm could be exploited in future studies of the AHM in lower dimensions (i.e., with layered/two-dimensional lattices) as well as for systems of ultracold atoms trapped in optical lattices.

\section*{Acknowledgments}
We thank G. Rohringer for several exchanges of ideas, and S. Ciuchi, K. Held, A. Katanin, F. B. Kugler, T. Schaefer, D. Springer, E. van Loon for insightful discussions. 
We acknowledge financial support from the Austrian Science Fund (FWF) through the projects I 2794-N35 and the SFB ViCoM F41 (AT), and the FSE-Friuli Venezia Giulia for the HEaD "Higher Education And Development" project FP1619889004. 
M.C. acknowledge support from  the H2020
Framework Programme  under ERC Advanced Grant No. 692670
``FIRSTORM', The Ministero dell'Istruzione Universit\`a e Ricerca through PRIN 2015 (Prot. 2015C5SEJJ001) and SISSA/CNR project "Superconductivity,
Ferroelectricity and Magnetism in bad metals" (Prot. 232/2015).

 \appendix
\section{Proof of the mapping} \label{Proof}

In this section, we prove that $\Sigma(k)$ as defined in Eqs.(\ref{Sigma:rep},\ref{Sigma:attr}) does not depend on the sign of the interactions at half filling, confirming the conclusions of the heuristic arguments presented in Sec.(\ref{sec:heuristic}).  

We divide the proof into three parts:
\begin{enumerate}
\item in the first, we derive some general properties relative to the mapping of  two-particle Green's functions of the exact solution under a partial particle-hole transformation, that will be used in the subsequent sections;
\item in the second part, we shall use a property of the 2PI local vertex function already derived in Ref.\cite{RohringerValli2012}, and show that the $F^{k,\,q-k^\prime,\, q-\Pi}_{pp}\leftrightarrow F_m^{k,\,k^\prime,\,-q}$ when $U\leftrightarrow-U$, where $\Pi = (\boldsymbol{\Pi},0)$, with $\boldsymbol{\Pi}\equiv(\underbrace{\pi,\pi,...,\pi}_d)$;
\item in the third part, we shall demonstrate that $F_{\uparrow\downarrow}^{k\,k^\prime,q} \equiv \frac{1}{2}\left(F^{k,k^\prime, q}_d-F^{k,k^\prime, q}_m\right)\leftrightarrow -F_{\uparrow\downarrow}^{k,\,-q-k^\prime,q} $, when $U \leftrightarrow -U$. To our knowledge this is an original result, since we did not find this relation in the literature.
\end{enumerate}
From the second and third points, it follows that Eq.({\ref{Sigma:attr}}) is mapped onto Eq.({\ref{Sigma:rep}}) when $U\to -U$.
\par Before entering the details of the proof, let us remind that the ladders $F_r$ are calculated via Bethe-Salpeter's equations, schematically depicted in Fig.(\ref{fig:bs}), that read:
\begin{eqnarray}\label{BS:ph}
F_{d/m}^{k,k^\prime,q} &=&  \Gamma_{d/m}^{\nu,\nu^\prime,\omega}+ \frac{1}{\mathcal{V}}\sum_{k_1}\Gamma_{d/m}^{\nu\nu_1\omega}\,G_{k_1}G_{k_1+q}\, F_{d/m}^{k_1,k^\prime,q} \nonumber \\
\\
\label{BS:pp}
F_{pp}^{k,k^\prime,q} &=&  \Gamma_{pp}^{\nu,\nu^\prime,\omega}- \frac{1}{\mathcal{V}}\sum_{k_1}\Gamma_{pp}^{\nu_1,\nu^\prime,\omega}\,G_{k_1}G_{q-k_1}\, F_{pp}^{k,q-k_1,q}, \nonumber \\
\end{eqnarray}
where $G_k = \left[i\nu+\mu-\Sigma(\nu)-\epsilon(\bk)\right]^{-1}$, with $\Sigma(\nu)$ being the local self-energy calculated with DMFT.
It is worth to notice that  the ladders depend on three frequency indices and only on the exchanged momentum $\mathbf{q}$, for simplicity we will keep the four-momentum notation for all the indices keeping in mind that $F_r(k,k^\prime,q)\equiv F_r(\nu,\nu^\prime,\omega;\mathbf{q})$.

\subsubsection{Transformation properties of two-particle Green's functions}
In this section we derive properties relative to the two-particle Green's function of the Hubbard model when the system posses particle-hole symmetry. In particular, we show how the two-particle Green's function transforms when a partial particle-hole transformation is performed.
\par The two particle Green's function is defined as:
\begin{equation}
G_{\sigma\sigma^\prime}(x_1,x_2,x_3,x_4)\equiv T_\tau\left<  \cd_{\sigma}(x_1)\cc_{\sigma}(x_2)\cd_{\sigma^\prime}(x_3)\cc_{\sigma^\prime}(x_4) \right>,
\end{equation}
where $x_i = (\bR_i,\tau_i)$, $c_\sigma(x_i)= e^{\tau_i \hat{H}}\,\cc_{\bR_i\sigma}\, e^{-\tau_i \hat{H}}$ and its Fourier components can be written as:
\begin{eqnarray}\label{G2:Fourier}
G_{\sigma\sigma^\prime}^{k,k^\prime,q}&\equiv& \int \prod_{i=1}^4 dx_i\, G_{\sigma\sigma^\prime}(x_1,x_2,x_3,x_4)\times\nonumber \\ 
&\times& e^{-i[kx_1-(k+q)x_2+(k^\prime+q)x_3-k^\prime x_4]},
\end{eqnarray}
where $k\,x = \tau\nu-\bk\cdot\bR$.
At half-filling, when a partial particle-hole transformation is performed, i.e. $\cc_{\bR\downarrow}\to e^{i \Pi\,x}\cd_{\bR\downarrow}$, $\hat{H}(U)\to \hat{H}(-U)$ and therefore, the two-particle Green's function is transformed in the following way $G_{(U),\uparrow\downarrow}(x_1,x_2,x_3,x_4) = -e^{-i\Pi(x_3-x_4)}G_{(-U),\uparrow\downarrow}(x_1,x_2,x_4,x_3)$, $G_{(U),\downarrow\downarrow}(x_1,x_2,x_3,x_4)= e^{-i\Pi(x_1-x_2+x_3-x_4)}G_{(-U),\downarrow\downarrow}(x_2,x_1,x_4,x_3)$. The system is $SU(2)$-symmetric, therefore $G_{\downarrow\downarrow}=G_{\uparrow\uparrow}$, and since the partial particle-hole transformation does not affect the spin-up fermions, i.e. $G_{(U),\uparrow\uparrow}(x_1,x_2,x_3,x_4) = G_{(-U),\uparrow\uparrow}(x_1,x_2,x_3,x_4)$, it is also true that $G_{(U),\uparrow\uparrow}(x_1,x_2,x_3,x_4)=e^{-i\Pi(x_1-x_2+x_3-x_4)}G_{(U),\uparrow\uparrow}(x_2,x_1,x_4,x_3)$.  Using these equalities in real space-time and Eq.(\ref{G2:Fourier}), we can write the following relations in Fourier space: $G_{(U),\uparrow\downarrow}^{k,k^\prime,q}=  -G_{(-U),\uparrow\downarrow}^{k,-q-k^\prime-\Pi,q}$, $G_{(U),\uparrow\uparrow}^{k,k^\prime,q} = G_{(U),\uparrow\uparrow}^{-q-k-\Pi,-q-k^\prime-\Pi,q}$.
It is easy to prove, from Eq.(\ref{G2:connected}), that these relations hold also for the 1PI-vertex function, namely:
\begin{eqnarray}\label{G2:mapping:updo}
\mathcal{F}_{(U),\uparrow\downarrow}^{k,k^\prime,q} &=&  -\mathcal{F}_{(-U),\uparrow\downarrow}^{k,-q-k^\prime-\Pi,q} , \\
\label{G2:mapping:upup}
\mathcal{F}_{(U),\uparrow\uparrow}^{k,k^\prime,q} &=&  \mathcal{F}_{(U),\uparrow\uparrow}^{-q-k-\Pi,-q-k^\prime-\Pi,q}.
\end{eqnarray}
Eq.(\ref{G2:mapping:updo}) implies that the exact self-energy of the Hubbard model at half-filling does not depend on the sign of the interactions, in fact $\Sigma_{(U)}(k) = U\sum_{k^\prime,q}G_{k+q}G_{k^\prime}G_{k^\prime +q}\mathcal{F}^{k,k^\prime,q}_{(U),\uparrow\downarrow} = (-U)\sum_{k^\prime,q}G_{k+q}G_{k^\prime}G_{k^\prime+q}\mathcal{F}_{(-U),\uparrow\downarrow}^{k,-q-k^\prime-\Pi,q} =(-U)\sum_{k^\prime,q}G_{k+q}G_{k^\prime}G_{k^\prime +q}\mathcal{F}^{k,k^\prime,q}_{(-U),\uparrow\downarrow}= \Sigma_{(-U)}(k) $, where for the last equality we performed a shift of the dummy indices, i.e. $k^\prime\to-q-k^\prime-\Pi$ and used the half-filling properties of the one particle Green's function $G(k) = -G(-k-\Pi)$ and $G_{(U)}(k)=G_{(-U)}(k)$.
\subsubsection{Relation between the particle-particle and magnetic ladders}
In Ref.\cite{RohringerValli2012}, it has been shown that at half-filling $\Gamma_{(U),m}^{\nu,\nu^\prime,-\omega} = \Gamma_{(-U),pp}^{\nu,\omega-\nu^\prime,\omega}$. If we substitute this into Eq.(\ref{BS:ph}) we obtain:
\begin{eqnarray}\label{BS:mapping1}
&&F_{(U),m}^{k,k^\prime,-q}=F_{(U),m}^{k^\prime,k,-q} =\nonumber \\
&=&\Gamma_{(-U),pp}^{\nu^\prime,\omega-\nu,\omega}+\frac{1}{\mathcal{V}}\sum_{k_1}\Gamma_{(-U),pp}^{\nu^\prime,\omega-\nu_1,\omega}G_{k_1}G_{k_1-q}F_{(U),m}^{k_1,k,-q} \nonumber \\
&=&\Gamma_{(-U),pp}^{\nu,\omega-\nu^\prime,\omega}-\frac{1}{\mathcal{V}}\sum_{k_1}\Gamma_{(-U),pp}^{\nu_1,\omega-\nu^\prime,\omega}G_{k_1}G_{q-\Pi-k_1}F_{(U),m}^{k,k_1,-q},  \nonumber \\
\end{eqnarray}
where we used time-reversal, space-inversion and $SU(2)$ symmetry, that together imply $F_{\sigma\sigma^\prime}^{k,k^\prime,q} = F_{\sigma\sigma^\prime}^{k^\prime,k,q}$ for both local and non-local quantities and in the second passage we used the following property of the Green's function: $G(k)=-G(k\pm\Pi)$ at half-filling.
If we perform the following shift $q \to q-\Pi$, $k^\prime \to q-k^\prime$ in Eq.(\ref{BS:mapping1}) and afterwards subtract the obtained expression from Eq.(\ref{BS:pp}), we  otbain the following homogenous system of linear equations for every couple of points $k$ and $q$:
\begin{equation}
\left(\mathbbm{1}+\mathbbm{A}_{pp}\cdot \mathbbm{S}\right)\cdot \mathbf{X} = \mathbf{0},
\end{equation}
where $\left[\mathbbm{A}_{pp}\right]_{k,k^\prime}=\frac{1}{\mathcal{V}}\,\Gamma_{(-U),pp}^{\nu,\nu^\prime,\omega}\,G_{k^\prime}G_{q-k^\prime}$, $[\mathbbm{S}]_{k,k^\prime} \equiv \delta_{k,q-k^\prime} $ and $\left[\mathbf{X}\right]_{k^\prime} \equiv  F_{(-U),pp}^{k,k^\prime,q}-F^{k,q-k^\prime,-q+\Pi}_{(U),m}$. If det$\left(\mathbbm{1}+\mathbbm{A}_{pp}\cdot \mathbbm{S}\right) \not = 0$ the only solution of the homogenous linear system is the trivial one, that reads explicitly:
\begin{equation}\label{ladder:mapping:mpp}
F^{k,\, k^\prime,q}_{(-U),\,pp} = F^{k,\,q-k^\prime,-q + \Pi}_{(U),m}
\end{equation}

Conversely, if non-trivial solutions are admitted, these solutions will not satisfy the mapping relation.
\subsubsection{Transformation of the $F_{\uparrow\downarrow}$.}
In the previous section we have shown that the property of the 1PI-vertex function in Eq.(\ref{G2:mapping:updo}) assures the right mapping of the self-energy at half-filling. This property holds for the exact case, therefore if  approximations are carried out it could no longer be valid. In this section, we prove that  this property that stems from particle-hole symmetry is preserved in the ladder approximation, namely:
\begin{equation}\label{ladder:mapping:updo}
 F^{k,k^\prime,q}_{(U),\uparrow\downarrow} = -F^{k,-q-k^\prime,q}_{(-U),\uparrow\downarrow}. 
 \end{equation}
 Eqs.(\ref{ladder:mapping:mpp},\ref{ladder:mapping:updo}) imply the right mapping of $\Sigma(k)$ that in our ladder approximation is defined in Eqs.(\ref{Sigma:rep},\ref{Sigma:attr}).
 Before addressing the proof, we shall express Eq.(\ref{BS:ph}) in alternative way that we found useful, that reads:
\begin{equation}\label{BS:ph1}
F_{d/m}^{k,k^\prime,q} =  \mathcal{F}_{d/m}^{\nu,\nu^\prime,\omega}+ \frac{1}{\mathcal{V}}\sum_{k_1} F_{d/m}^{k,k_1,q}\,\widetilde{G}_{k_1}\widetilde{G}_{k_1+q}\, \mathcal{F}_{d/m}^{\nu_1,\nu^\prime,\omega},
\end{equation}
where for brevity we used the notation $\mathcal{F}_{\sigma,\sigma^\prime}^{\nu,\nu^\prime,\omega} \equiv \mathcal{F}_{\sigma,\sigma^\prime}^{loc,\nu,\nu^\prime,\omega}$ that is the local 1PI-vertex function of the AIM, not to be confused with $\mathcal{F}_{\sigma\sigma^\prime}^{k,k^\prime,q}$, where the dependence on the momenta is fully taken into account, and $\widetilde{G}_k \equiv G_k - \frac{1}{V}\sum_\bk G(k)$ is the fully non-local Green's function that for construction has the following property $\sum_\bk\widetilde{G}_k =0$. First, let us note that $\mathcal{F}^{\nu\nu^\prime\omega}_{\sigma\sigma^\prime}$ is an exact quantity and therefore satisfies the relations in Eqs.(\ref{G2:mapping:updo} ,\ref{G2:mapping:upup}) at the local level.
Let us write explicitly the Bethe-Salpeter equations for the $F_{\uparrow\uparrow}$ and the $F_{\uparrow\downarrow}$ that can be obtained from Eq.(\ref{BS:ph1}):
\begin{eqnarray}\label{BS:ph2:eq1}
F^{k,k^\prime,q}_{\uparrow\uparrow} &=& \mathcal{F}^{\nu\nu^\prime\omega}_{\uparrow\uparrow} + \frac{1}{\mathcal{V}}\sum_{k_1}F^{kk_1q}_{\uparrow\uparrow}\widetilde{G}\widetilde{G}\mathcal{F}^{\nu_1\nu^\prime\omega}_{\uparrow\uparrow} \nonumber\\
&+& \frac{1}{\mathcal{V}}\sum_{k_1}F^{kk_1q}_{\uparrow\downarrow}\widetilde{G}\widetilde{G}\mathcal{F}^{\nu_1\nu^\prime\omega}_{\uparrow\downarrow}  \\
\label{BS:ph2:eq2}
F^{k,k^\prime,q}_{\uparrow\downarrow} &=& \mathcal{F}^{\nu\nu^\prime\omega}_{\uparrow\downarrow} + \frac{1}{\mathcal{V}}\sum_{k_1}F^{kk_1q}_{\uparrow\uparrow}\widetilde{G}\widetilde{G}\mathcal{F}^{\nu_1\nu^\prime\omega}_{\uparrow\downarrow} \nonumber\\
&+& \frac{1}{\mathcal{V}}\sum_{k_1}F^{kk_1q}_{\uparrow\downarrow}\widetilde{G}\widetilde{G}\mathcal{F}^{\nu_1\nu^\prime\omega}_{\uparrow\uparrow}, 
\end{eqnarray}
where $\widetilde{G}\widetilde{G}\equiv \widetilde{G}_{k_1}\widetilde{G}_{k_1+q}$.
Using the fact that $\mathcal{F}_{(U),\uparrow\uparrow}^{\nu\nu^\prime\omega} =\mathcal{F}_{(-U),\uparrow\uparrow}^{\nu\nu^\prime\omega}=\mathcal{F}_{(-U),\uparrow\uparrow}^{-\omega-\nu,-\omega-\nu^\prime,\omega} $ and that $\mathcal{F}_{(U),\uparrow\downarrow}^{\nu\nu^\prime\omega} =-\mathcal{F}_{(-U),\uparrow\downarrow}^{\nu,-\omega-\nu^\prime,\omega} =-\mathcal{F}_{(-U),\uparrow\downarrow}^{-\omega-\nu,\nu^\prime,\omega} $ 
we can manipulate Eqs.(\ref{BS:ph2:eq1},\ref{BS:ph2:eq2}) as following:
\begin{eqnarray}\label{Man:BS:eq1}
F^{k,k^\prime,q}_{(U),\uparrow\uparrow} &=& \mathcal{F}^{\nu\nu^\prime\omega}_{(-U)\uparrow\uparrow} + \frac{1}{\mathcal{V}}\sum_{k_1}F^{kk_1q}_{(U),\uparrow\uparrow}\widetilde{G}\widetilde{G}\mathcal{F}^{\nu_1\nu^\prime\omega}_{(-U),\uparrow\uparrow} \nonumber\\
&+& \frac{1}{\mathcal{V}}\sum_{k_1}(-F^{k,-q-k_1,q}_{(U)\uparrow\downarrow})\widetilde{G}\widetilde{G}\mathcal{F}^{\nu_1,\nu^\prime,\omega}_{(-U)\uparrow\downarrow}  \\
\label{Man:BS:eq2}
-F^{k,-q-k^\prime,q}_{(U),\uparrow\downarrow} &=& \mathcal{F}^{\nu,\nu^\prime,\omega}_{(-U),\uparrow\downarrow} + \frac{1}{\mathcal{V}}\sum_{k_1}F^{kk_1q}_{(U)\uparrow\uparrow}\widetilde{G}\widetilde{G}\mathcal{F}^{\nu_1,\nu^\prime,\omega}_{(-U)\uparrow\downarrow} \nonumber\\
&+& \frac{1}{\mathcal{V}}\sum_{k_1}(-F^{k,-q-k_1,q}_{(U)\uparrow\downarrow})\widetilde{G}\widetilde{G}\mathcal{F}^{\nu_1,\nu^\prime,\omega}_{(-U)\uparrow\uparrow}. 
\end{eqnarray}
If we subtract Eq.(\ref{Man:BS:eq1}) from Eq.(\ref{BS:ph2:eq1}) and Eq.(\ref{Man:BS:eq2}) from Eq.(\ref{BS:ph2:eq2}) we obtain the following homogenous system of equations for a generic couple of $q$ and $k$ points that is useful to write in a matrix notation:
\begin{equation}\label{sys:hom}
\left(
\begin{array}{cc}
\mathbbm{1}-\mathbbm{A}_{\uparrow\uparrow} & -\mathbbm{A}_{\uparrow\downarrow} \\
-\mathbbm{A}_{\uparrow\downarrow} &  \mathbbm{1}-\mathbbm{A}_{\uparrow\uparrow}
\end{array}
\right)
\left(
\begin{array}{c}
\mathbf{X} \\
\mathbf{Y}
\end{array}
\right) = \left(
\begin{array}{c}
\mathbf{0} \\
\mathbf{0}
\end{array}
\right),
\end{equation}

where $\left[\mathbbm{A}_{\sigma\sigma^\prime}\right]_{kk^\prime} \equiv \frac{1}{\mathcal{V}}\widetilde{G}_{k^\prime}\widetilde{G}_{k^\prime + q}\,\mathcal{F}_{(-U)\,\sigma\sigma^\prime}^{\nu, \nu^\prime,q} $, $\left[\mathbf{X}\right]_{k^\prime} \equiv F_{(-U)\uparrow\uparrow}^{k,k^\prime,q}- F_{(U)\uparrow\uparrow}^{k,k^\prime,q}$, $\left[\mathbf{Y}\right]_{k^\prime}\equiv F_{(-U)\uparrow\downarrow}^{k,k^\prime,q}-\left(-F^{k,-q-k^\prime,q}_{(U),\uparrow\downarrow} \right) $.
If the determinant of the matrix in Eq.(\ref{sys:hom}) does not vanish, the unique solution of the homogenous system is the trivial one and Eq.(\ref{ladder:mapping:updo}) is guaranteed. Conversely, if the homogenous system admits non-trivial solutions, these solutions would not satisfy the mapping relation in Eq.(\ref{ladder:mapping:updo}), likely leading to unphysical results.  We notice, however, that even in such a case the mapping properties of the linear combination defining  $F_m$  are preserved, due to the symmetry ($\mathbf{X}  \leftrightarrow \mathbf{Y}$)  of the Eqs.(\ref{sys:hom}).

\newpage

\bibliography{biblio} 

\end{document}